\documentclass[reprint, superscriptaddress]{revtex4-1}

\usepackage{amsmath}
\usepackage{graphicx}
\usepackage{float}
\usepackage[margin=1in]{geometry}

\usepackage{setspace}
\usepackage[sort&compress]{natbib}

\begin{document}

\author{Jack Hellerstedt}
\affiliation{School of Physics and Astronomy, Monash University, Victoria 3800, Australia}
\affiliation{Institute of Physics of the Czech Academy of Sciences, v.v.i., Cukrovarnick\'{a} 10, 162 00 Prague, Czech Republic}
\author{Indra Yudhistira}
\affiliation{Department of Physics and Centre for Advanced 2D Materials, National University of Singapore, 117551, Singapore}
\author{Mark T. Edmonds}
\author{Chang Liu}
\author{James Collins}
\affiliation{School of Physics and Astronomy, Monash University, Victoria 3800, Australia}
\affiliation{ARC Centre of Excellence in Future Low-Energy Electronics Technologies, Monash University, Victoria 3800, Australia}
\author{Shaffique Adam}
\affiliation{Yale-NUS College, 6 College Avenue East, 138614, Singapore}
\affiliation{Department of Physics and Centre for Advanced 2D Materials, National University of Singapore, 117551, Singapore}
\author{Michael S. Fuhrer}
\email{michael.fuhrer@monash.edu}
\affiliation{School of Physics and Astronomy, Monash University, Victoria 3800, Australia}
\affiliation{ARC Centre of Excellence in Future Low-Energy Electronics Technologies, Monash University, Victoria 3800, Australia}

\title{Electrostatic Modulation of the Electronic Properties of Dirac Semimetal Na$_3$Bi}

\begin{abstract}
Large-area thin films of topological Dirac semimetal Na$_3$Bi are grown on amorphous SiO$_2$:Si substrates to realise a field-effect transistor with the doped Si acting as back gate. As-grown films show charge carrier mobilities exceeding 7,000 cm$^2$/Vs and carrier densities below 3 $\times $10$^{18}$ cm$^{-3}$, comparable to the best thin-film Na$_3$Bi. An ambipolar field effect and minimum conductivity are observed, characteristic of Dirac electronic systems. The results are quantitatively understood within a model of disorder-induced charge inhomogeneity in topological Dirac semimetals. Due to the inverted band structure, the hole mobility is significantly larger than the electron mobility in Na$_3$Bi, and when present, these holes dominate the transport properties.
\end{abstract}

\maketitle

Topological Dirac semimetals (TDS) are three-dimensional analogues of graphene, with linear electronic dispersions in three dimensions \cite{Murakami2007, Young2012a, Wang2012i}. Thin films of TDS are of interest for topological devices; a conventional-to-topological quantum phase transition (QPT) occurs with increasing film thickness \cite{Wang2012i,Pan2015a,Xiao2015a}, and gate electrodes can enable an electric field-tuned QPT, realizing a topological transistor \cite{Pan2015a,Xiao2015a}. Additionally, nanostructured TDS exhibit unusual transport phenomena due to Fermi arc surface states \cite{Potter2014,Moll2015}. Field-effect structures are a key step towards realising topological devices based on TDS. To date field-effect gating has been realised on individual nanowires \cite{Zhang2015} or nanoplates \cite{Li2016c} of TDS Cd$_3$As$_2$, and electrolyte gating has been used on Cd$_3$As$_2$ thin films \cite{Liu2014b}. However no large-area solid-state process has been demonstrated for a TDS field-effect device.  

Here we report the growth of high quality thin films of Na$_3$Bi on amorphous SiO$_2$ on Si substrates. We have characterized the films using low temperature magneto-transport \emph{in situ} in ultrahigh vacuum (UHV) \cite{Hellerstedt2014, Hellerstedt2016, Edmonds2016}. As-grown films show charge carrier mobilities exceeding 7,000 cm$^2$/Vs and carrier densities below 3 $\times $10$^{18}$ cm$^{-3}$, comparable to the best thin-films on Al$_2$O$_3$ \cite{Hellerstedt2016, Edmonds2016}. Using the doped Si substrate as a back-gate electrode, the carrier density can be tuned over ~2 $\times $10$^{18}$ cm$^{-3}$. Using a combination of molecular surface transfer doping \cite{Edmonds2016} and field-effect gating, an ambipolar field effect and minimum conductivity are observed, characteristic of Dirac electronic systems near the Dirac point. The results are quantitatively understood within a model of disorder-induced charge inhomogeneity in topological Dirac semimetals. The hole mobility is found to be significantly higher than the electron mobility in Na$_3$Bi, reflecting the unusual character of the inverted band structure.

\begin{figure*}
\centering
\includegraphics[keepaspectratio=true, width=1\linewidth]{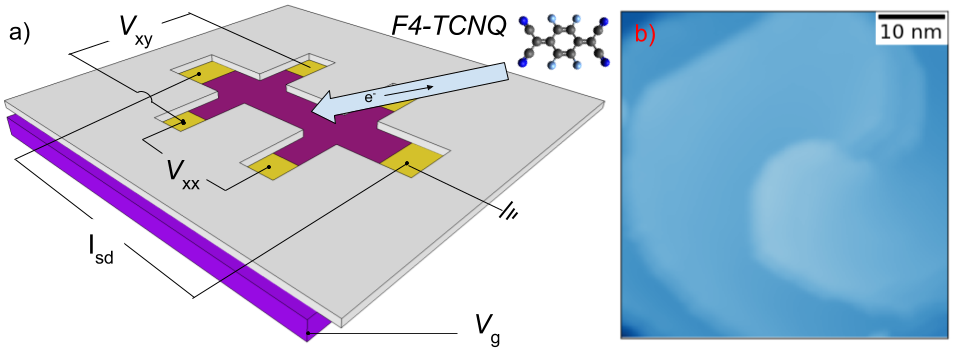}
\caption{a) Device schematic, showing Hall bar film geometry defined by the surface stencil mask affixed on the SiO$_2$ on Si substrate that serves as a back gate. After growth of the Na$_3$Bi thin film, the molecular charge acceptor F4-TCNQ is subsequently deposited on the top surface.  b) Scanning tunneling microscopy (STM) morphology of sample. Scan area is 50 $\times$ 50 nm, vertical color scale is 2 nm, STM bias voltage V$_{bias}$ = -300 mV, and current I = 200 pA.  The observable step edges have height of 4.2\AA, close to the expected 4.83\AA \, half-unit cell \cite{Wang2012i}.}
\label{sample-fig}
\end{figure*}

A Na$_3$Bi thin-film of thickness 18 nm was deposited through a stencil mask onto a pre-existing electrode pattern on SiO$_2$ (1 $\mu$m) on Si substrates \cite{supplemental}. Figure \ref{sample-fig}a) shows a schematic of the sample and stencil mask used to define the Hall bar geometry. After growth of the thin film on the pre-existing electrical contacts, the entire structure is transferred within UHV to the analysis chamber where electrical measurements are performed at a temperature of 5 K, and scanning tunneling microscopy is also performed. Figure \ref{sample-fig}b) shows a scanning tunneling microscopy image of the film surface. The film is highly oriented with $c$-axis perpendicular to the substrate, showing atomically flat [0001] terraces of width 20 nm or more. Since the as-grown films are found to be $n$-type, we employ additional $p$-type doping with the organic molecule 2,3,5,6-Tetrafluoro-7,7,8,8-tetracyanoquinodimethane (F4-TCNQ). F4-TCNQ acts as a \emph{p}-type dopant due to its high electron affinity, and physisorbs on the surface of Na$_3$Bi with little detrimental effect on the charge carrier mobility \cite{Edmonds2016}. The Na$_3$Bi film thickness of 18 nm is comparable to or less than the Thomas-Fermi screening length at the observed charge carrier densities, indicating that band-bending across the thickness of the film is not significant and hence field-effect and molecular doping should affect the entire film thickness.

\begin{figure}
\centering
\includegraphics[keepaspectratio=true, width=1\linewidth]{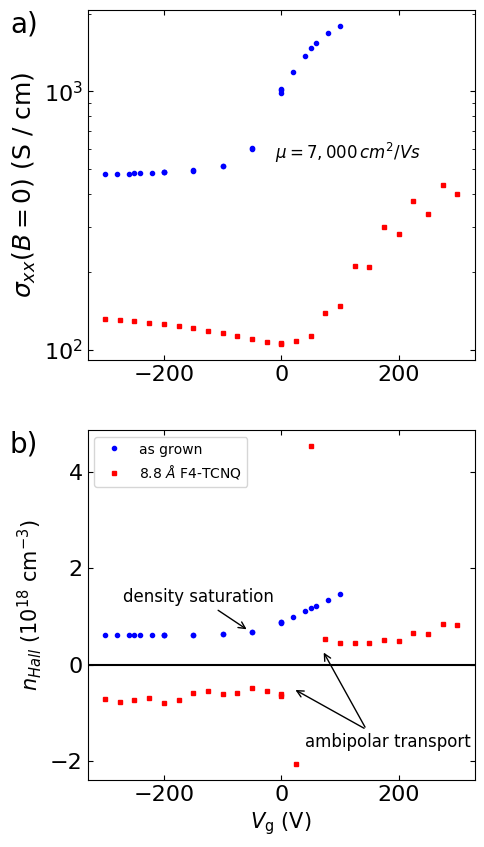}
\caption{a) Zero field conductivity (S/cm) and b) low-field Hall carrier density (10$^{18}$ cm$^{-3}$) vs. back gate voltage (V$_g$), for the Na$_3$Bi film as-grown (blue circles), and after the deposition of 8.8\AA \, F4-TCNQ on the surface (solid squares).}
\label{as-grown-sigma-fig}
\end{figure}

Figure \ref{as-grown-sigma-fig} shows the conductivity $\sigma_{xx}$ (Fig. \ref{as-grown-sigma-fig}a) and Hall carrier density $n_{Hall}$ (Fig. \ref{as-grown-sigma-fig}b) as a function of gate voltage $V_{\mathrm{g}}$ applied to the silicon, for Na$_3$Bi films as-grown and after doping with 8.8 \AA \, of F4-TCNQ. The Hall carrier density is calculated from the low-field ($B < .05$ T), linear Hall response: $n_{Hall} = B/(e \, \rho_{xy})$. The as-grown Na$_3$Bi film (blue circles) has an $n$-type carrier density of 8.8 $\times$ 10$^{17}$ cm$^{-3}$ and a conductivity 1.0 $\times$ 10$^{3}$ S/cm at $V_{\mathrm{g}}= 0$ V. This corresponds to a Hall mobility of 7,200 cm$^{2}$/Vs.  The conductivity decreases with decreasing Hall carrier density as expected for the $n$-type film. The gate dependent carrier density for the as-grown film in b) shows non-linear saturation for negative gate voltages, probably due to charge trapping in the oxide due to degradation by the Na/Bi deposition. 

The same Na$_3$Bi film with an 8.8 \AA \, overlayer of F4-TCNQ shows qualitatively different behaviour (red squares). The conductivity of the sample is reduced by approximately a factor of four, and the gate dependent conductivity reveals a minimum. The Hall carrier density shows an asymptotic positive (negative) divergence in carrier density at gate voltages roughly above (below) the gate voltage of minimum conductivity. Qualitatively similar behaviour is observed in other gapless Dirac semimetals such as graphene \cite{Novoselov2004} and the surface state of topological insulator Bi$_2$Se$_3$ \cite{Kim2012}, where the minimum conductivity and the change of sign of Hall carrier density reflect the Fermi energy crossing the Dirac point.

\begin{figure*}
\centering
\includegraphics[keepaspectratio=true, width=1\linewidth]{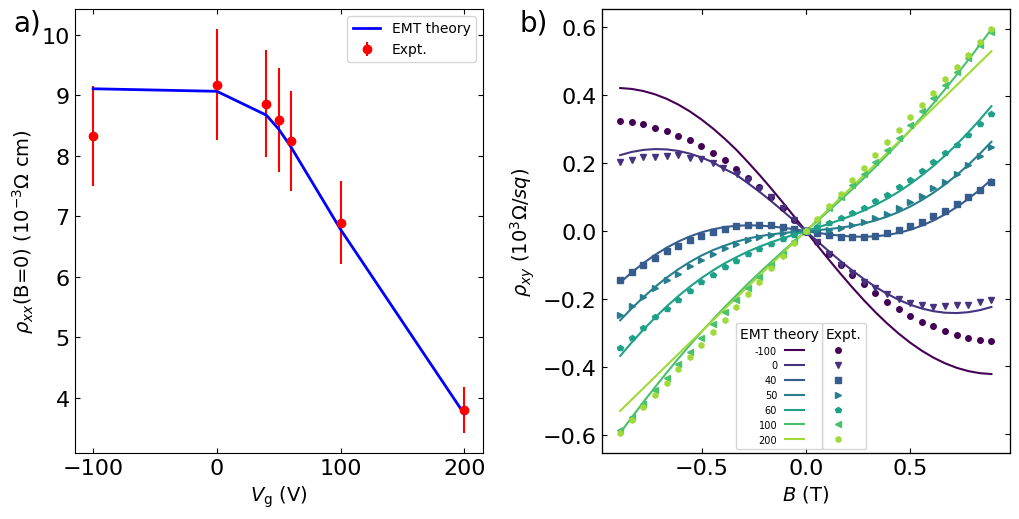}
\caption{a) Gate dependence of the zero-field resistivity $\rho_{xx}$ ($\Omega$-cm) after top-gating with F4-TCNQ.  The red points are the measured values; the blue line is the theoretical fit.  b) Transverse magnetoresistance $\rho_{xy}$ as a function of magnetic field perpendicular to the film, for back gate voltages (V$_g$) indicated in the legend.  The markers are the measured data, while the solid lines are fits using the charge inhomogeneity theory described in the text.}
\label{MR-data-fig}
\end{figure*}

To better understand the physics close to the crossing of the Dirac point, in Fig. \ref{MR-data-fig} we show the detailed longitudinal and Hall resistivity of the F4-TCNQ-doped sample in the region close to the sign change in the carrier density.  Figure \ref{MR-data-fig}a) shows the gate-voltage dependence of the zero-field longitudinal resistivity $\rho_{xx}$, showing a maximum at V$_g$ = 0 V. Figure \ref{MR-data-fig}b) shows the magnetic field dependence of the Hall resistivity $\rho_{xy}$ at various gate voltages. There are several notable features in the gate dependent Hall response. At low field ($B <$ 0.25 T) there is a clear transition from electron- to hole-like response as the gate is modulated from positive to negative bias; this results in the sign change of $n_{Hall}$ observed in Fig. \ref{as-grown-sigma-fig}b). The smooth change of slope from positive to negative through zero results in the divergence of $n_{Hall}$ observed in Fig. \ref{as-grown-sigma-fig}b), as $n_{Hall}$ is inversely proportional to the slope. Furthermore, as the low-field response tends to hole-like (negative slope), the overall signal becomes non-linear, which indicates contributions from multiple carrier types \cite{Yang2014, Bansal2011, Ali2014}. Note that the as-grown Na$_3$Bi sample showed completely linear $\rho_{xy}(B)$ over the entire range of gate voltage, consistent with a single $n$-type carrier.

We first attempted to fit the $\rho_{xy}(B)$ data using a two-carrier model. The change in slope of $\rho_{xy}(B)$ with $B$ requires a hole channel of lower concentration and higher mobility than the electron channel, and individual $\rho_{xy}(B)$ curves can be well fit to the two carrier model. However, the implied gate dependence of the total carrier number (e.g. the sum of the concentrations of the two channels, $n$ + $p$) is unphysical, indicating e.g. that the number of electrons in the sample is increasing with more negative applied gate voltage.

Failure of the two-carrier model is not surprising. Transport near the Dirac point is understood to involve spatially inhomogeneous regions of different carrier sign and density, with qualitatively different behaviour than the two-carrier model \cite{Adam2007}. Therefore to quantitatively understand our results, we have applied the previously developed effective medium theory (EMT) for charge inhomogeneity in graphene \cite{Adam2007}, modified to apply to Weyl semimetals for the relevant field regime ($\mu B<1$) \cite{Ramakrishnan2015}.  For the data taken following the F4-TCNQ deposition, we can fit the data in Fig. \ref{MR-data-fig}a) \& \ref{MR-data-fig}b) at each applied gate voltage using three global fit parameters. Two parameters $A_e$, $A_h$ are proportional to the electron and hole mobilities respectively; they are defined as the constants of proportionality between $\sigma_{xx}$ and $|n|^{4/3}$ (the expected carrier density dependence for a TDS \cite{DasSarma2015}) and depend on the impurity density $n_{imp}$, the electron and hole Fermi velocities $\nu_e$ and $\nu_h$, and the dielectric constant $\kappa$. A third global parameter, $n_{rms}$, is the measure of the disorder induced carrier density fluctuations \cite{Ping2014}. We take the carrier density $n$ as the only V$_g$-dependent fit parameter.

\begin{figure}
\centering
\includegraphics[keepaspectratio=true, width=1\linewidth]{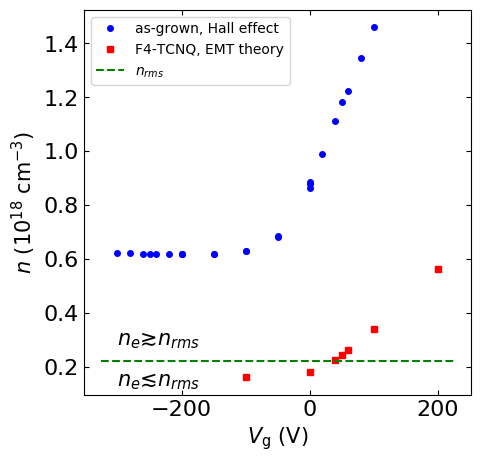}
\caption{Carrier density $n$ as a function of gate voltage $V_{\mathrm{g}}$ for the as-grown sample as determined from Hall effect (blue circles) and for the F4-TCNQ-doped sample from the EMT fits in Fig. \ref{MR-data-fig}b) (red squares). The solid green line is the value of the global fit parameter $n_{rms}$ = 2.2 $\times$ 10$^{17}$ cm$^{-3}$.  The blue circles are the as-grown carrier density for comparison.}
\label{n-fit-fig}
\end{figure} 

Solid lines in Fig. \ref{MR-data-fig} show the results of the fits to the EMT. The EMT fits reproduce the very shallow maximum in $\rho_{xx}(V_{\mathrm{g}})$, and the non-linear $\rho_{xy}(B)$. As anticipated, we find a larger hole mobility than electron mobility, as reflected in the parameters A$_h$ = 1.5 $\times 10^{-1}$ and A$_e$ = 1.4 $\times 10^{-2}$ $e^2$/ $h$ / 10$^{15}$ cm$^{-3}$. The final global fit parameter is the Gaussian width that determines the size of the charge inhomogeneous regime $n_{rms}$ = 2.2 $\times$ 10$^{17}$ cm$^{-3}$. 

Figure \ref{n-fit-fig} shows the carrier density extracted from the fits as a function of gate voltage, as well as the measured Hall carrier density for the film before F4-TCNQ doping, expected to be in the homogeneous regime. The overall dependence of the net carrier density $n$ on $V_{\mathrm{g}}>0$ for the F4-TCNQ doped film determined from fitting (red squares) and as-grown (blue circles) has similar slope in all samples we measured (see supplementary materials for details) \cite{supplemental}.  Surprisingly, the net carrier density at all gate voltages remains $n$-type even after F4-TCNQ doping. The implication is that the change in sign of the Hall carrier density (Fig. \ref{as-grown-sigma-fig}b) results from minority holes which dominate the Hall signal due to their much higher mobility. The value of $n_{rms}$ = 2.2 $\times$ 10$^{17}$ cm$^{-3}$, is also indicated for reference (green dashed line). As the net carrier density is reduced and approaches this inhomogeneous regime, the $n$-type fraction of the sample decreases while the $p$-type fraction increases. The conductivity minimum at $V_{\mathrm{g}}$ = 0 V (Fig. \ref{as-grown-sigma-fig}a) thus does not correspond to a change in carrier sign; this counterintuitive observation is not unexpected in the case of this huge asymmetry between electron and hole conductivities \cite{Adam2012}.

Analysis of the magnetoresistance (MR) $\rho_{xx}(B)$ is complicated by the large weak anti-localisation (WAL) contribution in our thin films, which is not captured by the semi-classical EMT.  We have attempted to subtract the WAL component in order to compare the experimental MR to the EMT.  We find that EMT still overestimates the magnitude of the WAL-corrected quadratic low-field MR.  However, anomalies in high field measurements of the resistivity tensor of Na$_3$Bi (including $\vec{E} \perp \vec{B}$ comparable to the present geometry) have been previously noted and attributed to a $B$-dependent transport lifetime $\tau_{tr}$ \cite{Xiong2015}.  If the $\tau_{tr}(B)$ dependence is identical for electrons and holes, only the $\rho_{xx}$ component is affected, explaining why we achieve good agreement in fitting only the $\rho_{xy}$ component.  MR data and further detail is provided in the supplementary materials \cite{supplemental}.

The magnitude of the charge inhomogeneity $n_{rms}$ can be used to estimate the potential fluctuations in the sample due to charge puddling, $E_{rms}$. Taking the Fermi velocity to be in the range $v_F = 1.4 - 2.4 \times 10^5$ m/s \cite{Hellerstedt2016, Liu2014a} we obtain $E_{rms} = 8.9-15$ meV and 4-6.9 meV for the inhomogeneous ($n \ll n_{rms}$) and homogeneous ($n \gg n_{rms}$) limits respectively; we expect that the actual value lies between these limits as $n \sim n_{rms}$. These values are 1-3 times larger than measured from direct measurements of the charge puddles \cite{Edmonds2017}, but this is not surprising because the amorphous SiO$_2$ substrate used in this case should introduce considerably more structural disorder than the atomically flat substrates used in the STM study. 

The hole conductivity coefficient A$_h$ is an order of magnitude larger than the electron coefficient A$_e$, implying a hole mobility an order of magnitude higher than electron mobility. This is already implied by the non-linear $\rho_{xy}(B)$: at low field, $\rho_{xy}$ is the mobility-weighted average and is dominated by the high-mobility minority holes, and the slope is negative ($p$-type). At fields higher than the inverse mobility of the holes, the Hall angle for the holes approaches $\pi/2$, and the lower mobility but higher concentration electrons begin to dominate producing a tendency to positive slope ($n$-type). Higher mobility for holes compared to electrons is unusual in conventional semiconductors. However the TDS band structure results from a band inversion, in which the shallower maximum of the more massive $p$-band occurs above the Fermi energy, and the deeper minimum of the less massive $s$-band occurs below the Fermi energy. This naturally results in a situation where the hole velocity exceeds the electron velocity \cite{Wang2012i}.

We note that temperature dependent transport measurements on bulk single crystal Na$_3$Bi have shown the dominance of hole-like transport above 140K \cite{Xiong2015b}. It is possible in that case that thermal activation of high-mobility holes causes a $p$-type Hall signal in net electron-doped crystals, similar to the effect of inhomogeneity observed here. 

In summary, we have demonstrated the growth of high quality, large area thin films of Na$_3$Bi directly on amorphous SiO$_2$ on Si realising a field-effect transistor structure, a significant step towards TDS devices.  The gate voltage-dependent conductivity shows a non-zero minimum, and the Hall carrier density shows a transition from electron-like to hole-like, and a divergence to positive (negative) for gate voltages above (below) the minimum conductivity. The gate voltage- and magnetic field-dependent longitudinal and Hall conductivity are well described with the recently developed theory of charge inhomogeneity in a Dirac semimetal system using physically realistic parameters \cite{Ramakrishnan2015}. The modeling indicates that holes have higher mobility than electrons, a natural consequence of the inverted band structure of Na$_3$Bi. As a result, the minimum conductivity and sign change of the Hall effect occur when the sample is still net $n$-type.

This work was performed in part at the Melbourne Centre for Nanofabrication (MCN) in the Victorian Node of the Australian National Fabrication Facility (ANFF).  I.Y. and S.A. are supported by the National Research Foundation of Singapore under its Fellowship program (NRF-NRFF2012-01).  J.H., M.T.E., C.L., J.C., S.A. and M.S.F. are supported by the ARC under awards FL120100038 (J.H., M.T.E., M.S.F), CE170100039 (M.T.E, C.L., J.C., M.S.F.), DE160101157 (M.T.E) and DP150103837 (S.A., M.S.F.). J.H. acknowledges the support of the CIES/ Czech Fulbright commission during the data analysis and preparation of this manuscript.



\setlength{\bibsep}{0pt}
\bibliography{bib-file}

\end{document}